\documentclass[conference]{IEEEtran}

\IEEEoverridecommandlockouts
\usepackage{cite}
\usepackage{amsmath,amssymb,amsfonts}
\usepackage{algorithmic}
\usepackage{graphicx}
\usepackage{mathptmx}
\usepackage{textcomp}
\usepackage{xcolor}
\usepackage{subfigure}
\usepackage{epstopdf}
\usepackage{booktabs}
\usepackage{epstopdf}
\usepackage{blindtext}
\usepackage{diagbox}
\def\BibTeX{{\rm B\kern-.05em{\sc i\kern-.025em b}\kern-.08em
    T\kern-.1667em\lower.7ex\hbox{E}\kern-.125emX}}

\begin{document}

\title{Design and Process Analysis of a Split-Gate Trench Power MOSFET with Bottom-Trench High-$k$ Pillar Superjunction for Enhanced Performance



\author{
    Yunteng Jiang$^{1,2}$, Zhentao Xiao$^{1,3}$, Zonghao Zhang$^{1,3}$, Juncheng Zhang$^{2}$, Chenxing Wang$^{1,3}$, \\
    Wenjun Li$^{1}$, Haimeng Huang$^{1*}$, Aynul Islam$^{1,3}$, Hongqiang Yang$^{1}$ \\
    \\
   \small \textit{$^1$University of Electronic Science and Technology of China, Chengdu 610054, China} \\
    \small \textit{$^2$Nanyang Technological University, Singapore 639798, Singapore} \\
    \small \textit{$^3$University of Glasgow, Glasgow, G12 8QQ, United Kingdom} \\
    *Corresponding author: hmhuang@uestc.edu.cn
}
}

\maketitle

\begin{abstract}

In this paper, we propose a simulation-based novel Split-Gate Trench MOSFET structure with an optimized fabrication process to enhance power efficiency, switching speed, and thermal stability for high-performance semiconductor applications. Integrating High-$k$ pillars Superjunction beneath the Split-Gate  enhancing breakdown performance by reducing critical field intensity by up to 35\%, the device achieves a 15\% improvement in Figures of Merit (FOMs) for $\rm{BV}^{2}/R_{\rm{on,sp}}$. Dynamic testing reveals approximately a 25\% reduction in both input and output capacitance, as well as gate-drain charge ($Q_{\text{GD}}$). This reduction, coupled with an approximately 40\% improvement in Baliga’s High-Frequency Figure of Merit (BHFFOM) and over 20\% increase in the New High-Frequency Figure of Merit (NHFFOM), underscores the design’s suitability for high-speed, high-efficiency power electronics. Simulations examining the effects of High-$k$ pillar depth indicate that an optimal depth of 3.5 $\mu$m achieves a balanced performance between BV and $R_{\text{on,sp}}$. The influence of High-$k$ materials on BT-H$k$-SJ MOSFET performance was investigated by comparing hafnium dioxide (HfO$_2$), nitride, and oxynitride. Among these, HfO$_2$ demonstrated optimal performance across static, dynamic, and diode characteristics due to its high dielectric constant, while material choice had minimal impact, with variations kept within 5\%. 
\par\

\end{abstract}


\section{Introduction}
\par Trench MOSFETs, as advancements over VD-MOSFETs, effectively eliminate the JFET effect, reduce conduction losses, and enhance current-handling capabilities, making them widely used in modern power electronics \cite{TrenchZS}\cite{BaligaB.Jayant2008Fops}. Their compact design and well-established fabrication processes make them ideal for a variety of power applications. To meet the demands of high-frequency operations, continuous improvements in key Figures of Merit (FOM) are essential for Trench MOSFETs. One significant enhancement is the Split-Gate design, which notably reduces Miller plateau capacitance and switching charge, thereby improving switching performance—crucial for high-frequency applications\cite{TongC.F.2009SaDA}. By minimizing Miller capacitance, the Split-Gate Trench MOSFET (SGT-MOSFET) achieves faster switching speeds and greater overall efficiency, making it particularly suitable for high-performance power devices.

\par Despite these advantages, there remains room for further optimization of the SGT structure. Current Trench MOSFETs face limitations in manufacturability, parasitic resistance, high gate-drain capacitance, and reliability. These issues complicate high-volume production, reduce efficiency in high-speed applications, highlighting the need for design improvements\cite{HistoryofTRENCH}. In industrial applications, the trade-off between breakdown voltage ($\text{BV}$) and specific on-resistance ($R_{\text{on,sp}}$), alongside the need to minimize conduction and switching losses, is crucial. Therefore, our objective is to enhance the performance of SGT-MOSFETs by addressing these specific challenges and improving their design for greater efficiency.

\begin{figure}
    \centering
\includegraphics[width=0.8\columnwidth]{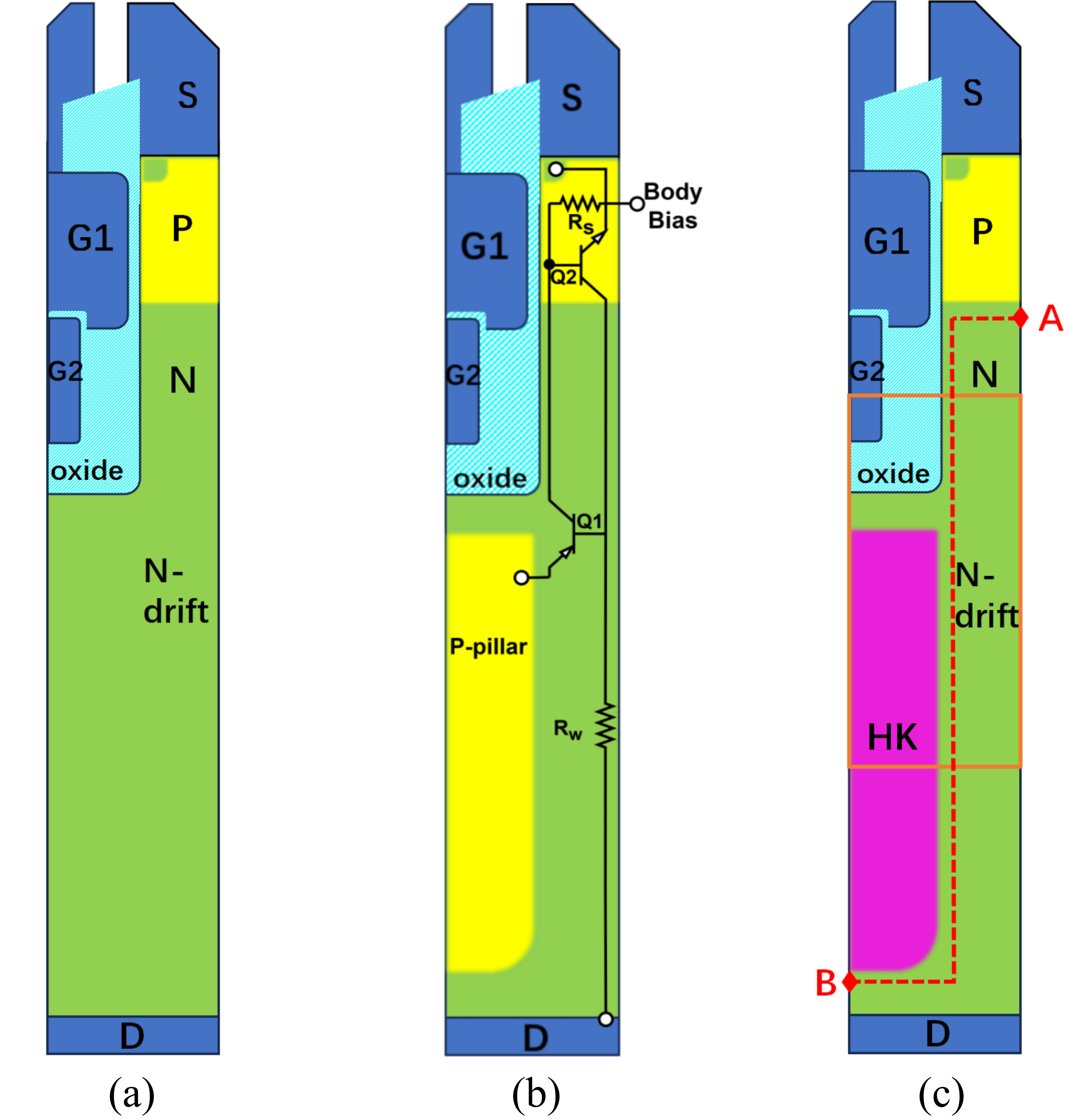}
     \caption{Vertical electric field and three dimensional electric field distribution of BT-H$k$-SJ MOSFET and conventional SGT-MOSFET}
    
\end{figure}

\par Typically, the space beneath the trench in SGT structures is underutilized, as shown in Fig. 1(a), with minimal current flow and no contribution to $\text{BV}$. This represents an opportunity for structural optimization in the unused region below the trench. Inspired by the Superjunction (SJ) structure, which enhances breakdown voltage through precise charge compensation and reduces $R_{\text{on,sp}}$ via alternating highly-doped $n$- and $p$-pillars\cite{udrea2017superjunction,chen2000novel}, we aim to incorporate these mechanisms into the SGT-MOSFET design.

\par Integrating SJ in the under-trench region of SGT-MOSFETs can potentially improve breakdown voltage and increase electric field concentration in the drift region due to the depletion effect\cite{HE201958}. However, achieving a conventional SJ approach by fabricating an under-trench $p$-pillar presents significant challenges, particularly in maintaining precise charge balance and controlling dopant diffusion in a desire shape within the device\cite{udrea2017superjunction,HkSJ1}. Moreover, the presence of the $p$-pillar can introduce parasitic BJT effects ,illustrated in Fig. 1(b), which may lead to latch-up and subsequent premature breakdown of the device.


\par To overcome these challenges, we propose using Bottom-Trench High-$k$ pillars Superjucntion (BT-H$k$-SJ) as an alternative ,shown in Fig 1.(c). These structure facilitates effective charge balancing by forming a High-$k$ SJ structure. This also allows higher doping concentrations in the $n$-pillar, eliminating the need for a $p$-pillar, and enhances $\text{BV}$ without compromising $R_{\rm{on,sp}}$. This could result in simplified fabrication, superior dynamic and breakdown performance\cite{HkSJ1,HkSJ2, zzh}. Additionally, the absence of parasitic BJTs helps to prevent latch-up and premature breakdown. In general,  BT-H$k$-SJ MOSFET is expected to achieve higher breakdown voltage and improved power density with reduced fabrication complexities.

\par In this study, we first present the fabrication process for our novel BT-H$k$-SJ MOSFET, emphasizing process viability for integrating and fabricating High-$k$ pillars in deep trenches using TSUPREME4 and MEDICI simulation. Next, we provide a comprehensive analysis of on the device structure and device performance, through various deep trench depths for SJ High-$k$ pillars with optimized processes, showcasing a comprehensive examination of both static and dynamic performance. We also investigate the High-$k$ pillar deposition using different materials, demonstrating the adaptability of the proposed fabrication method and the tunability of performance through material selection.

\section{Brief Fabracation Process}

\begin{figure}[htbp]
    \centering
    \subfigure[]{
        \includegraphics[width=0.155\columnwidth]{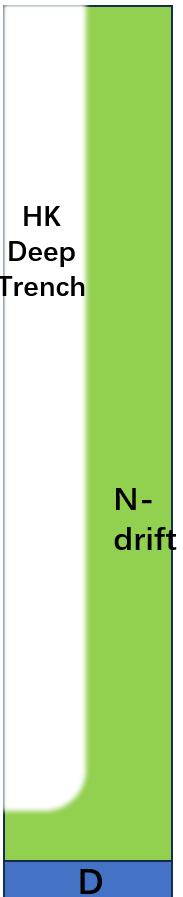}
        
    }\hspace{0.01\linewidth} 
    \subfigure[]{
        \includegraphics[width=0.16\columnwidth]{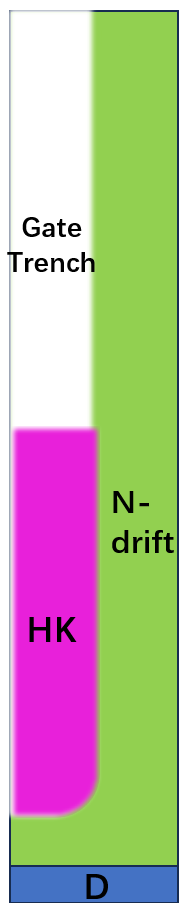}
        
    }\hspace{0.01\linewidth} 
    \subfigure[]{
        \includegraphics[width=0.16\columnwidth]{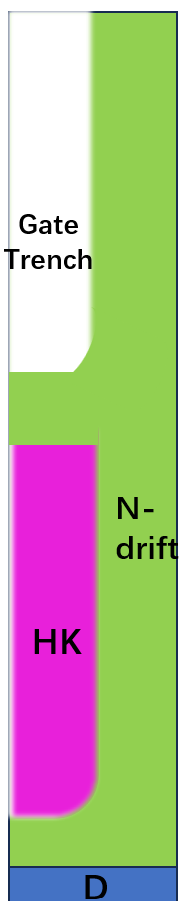}
        
    }
    \subfigure[]{
        \includegraphics[width=0.15\columnwidth]{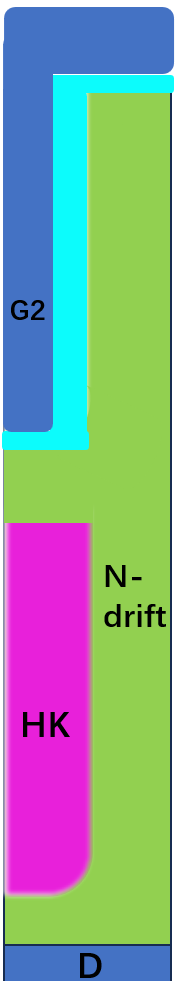}
        
    }

    \centering
    \subfigure[]{
        \includegraphics[width=0.155\columnwidth]{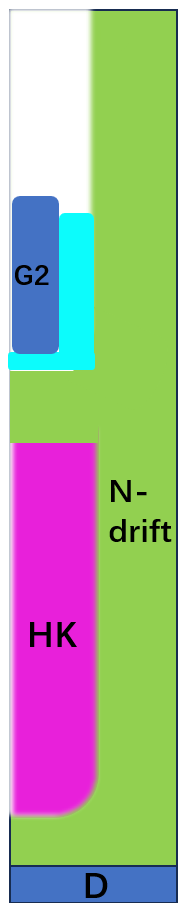}
        
    }\hspace{0.01\linewidth} 
    \subfigure[]{
        \includegraphics[width=0.16\columnwidth]{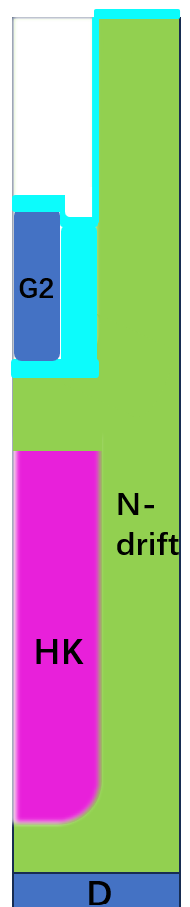}
        
    }\hspace{0.01\linewidth} 
    \subfigure[]{
        \includegraphics[width=0.15\columnwidth]{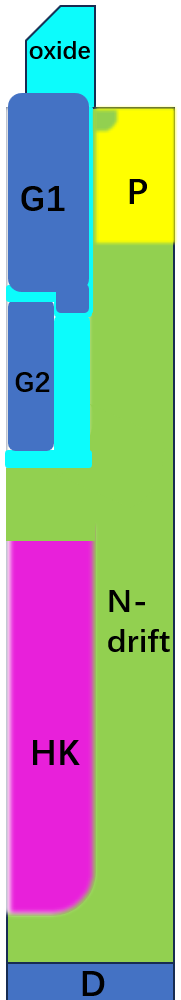}
        
    }
    \subfigure[]{
        \includegraphics[width=0.187\columnwidth]{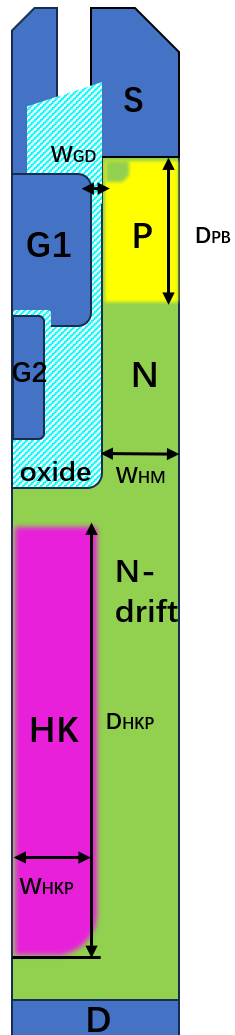}
        
    }
    \caption{Key process steps to fabricate SGT }
    
\end{figure}

\par The fabrication process of purposed BT-H$k$-SJ MOSFET is simulated and verified via TSUPREME4. The process begins with the growth of a lightly doped silicon layer on a highly doped silicon substrate ($N_{\rm{D}}=10^{17}$cm$^{-3}$) to serve as the drift region. A deep trench is then etched at a designated location using reactive ion etching (RIE) \cite{DoubleDiffuse}, demonstrated in Fig. 2(a). Then, this trench is filled with a High-$k$ dielectric material through magnetron sputtering deposition, depicted in Fig. 2(b), followed by a silicon deposition process on the High-$k$ pillar, illustrated in Fig. 2(c). Subsequently, oxide and polysilicon layers are deposited in sequence and etched into suitable shapes, as shown in Fig. 2(d). A thin and dense gate oxide is formed through thermal oxidation, represented in Fig. 2(e). Afterward, the control gate is deposited, and the $p$-body and $n^{+}$ region are created through ion implantation and high-temperature annealing, as illustrated in Fig. 2(f). Finally, metal is deposited on the source and gate for Ohmic contact, followed by etching for desired pattern. These process are used to fabricate the novel power MOSFET proposed in this paper, achieving an Botton-Trench High-$k$ Pillar with the shape of High-$k$ Superjunction structures which is demonstrated in Fig. 2(h). The dimensions of the structure and the doping parameters are given in Table I.

\begin{table}[htbp]
\centering
\caption{\textbf{Used Parameters in Simulation}}
\begin{tabular}{ll}
\toprule
Symbol & Value \\ \midrule
$W_{\text{HM}}$ (Half mesa width, $\mu$m) & 0.5 \\
$W_{\text{GD}}$ (Gate dielectric width, $\mu$m) & Optimized \\
$W_{\text{HKP}}$ (High-$k$ pillar width, $\mu$m) & 0.5 \\
$D_{\text{HKP}}$ (High-$k$ pillar depth, $\mu$m) & 2-4.5 \\
$p$-body depth ($\mu$m) & 0.46 \\
$N_{\text{D}}$ (Drift region concentration, cm$^{-3}$) & Optimized \\
$N_{\text{PD}}$ ($p$-body concentration, cm$^{-3}$) & Optimized \\
\bottomrule
\end{tabular}
\end{table}



\begin{figure}[htbp]
    \centering

        \includegraphics[width=1\columnwidth]{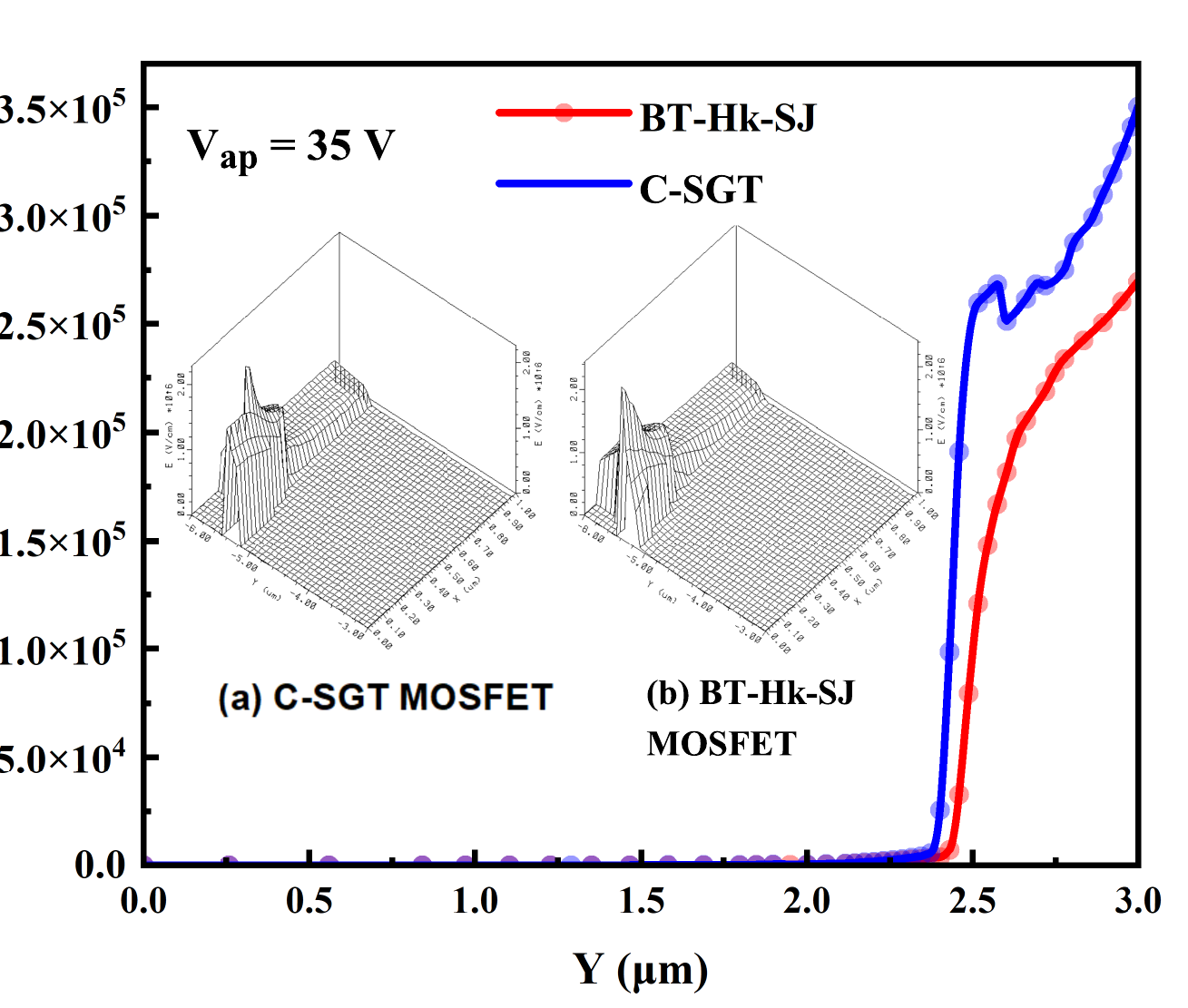}
       
    \hspace{0.01\linewidth} 
   \setlength{\abovecaptionskip}{0.cm}
      \vspace{-6mm}
      
     \caption{Vertical electric field and three dimensional electric field distribution of BT-H$k$-SJ MOSFET and conventional SGT-MOSFET}
    
\end{figure}

\section{Device Structural and Mechanism Analysis}

\par In order to precisely analyze and verify our device performance based on our proposed structure and fabrication, a comprehensive MEDICI simulation is performed. In the MEDICI device simulation process, several models are specifically incorporated, including AN-ALYTIC, FLDMOB, conSRH, IN-COMPLETE, IMPACT. I, ENERGY.L, BGN and FERMIDIR. All characteristics are tested when the drain-source on-state resistance ($R_{\rm{DS}}$) and threshold voltage ($V_{\rm{th}}$) are set to 0.74 m$\Omega$ ($V_{\rm{GS}}$=4.5V) and 1.5V at room temperature respectively.

\subsection{\textbf{Breakdown Mechanism Analysis}}

\par  Incorporating a High-$k$ insulator pillar can increase the effective permittivity in the region under the Split-Gate, leading to the absorption electric field previously distributed in $n$-drift region, which decreases the electric field component along the impact ionization path such as AB in Fig. 1(c). Considering avalanche breakdown, the reduction of critical breakdown field in the drift region is expected to increase the $\text{BV}$ of the BT-H$k$-SJ MOSFET, enhancing its resilience to doping variations and optimizing $R_{\rm{on,sp}}$ \cite{Mingmin_2016}. The vertical electric field and 3-D electric field distribution of the BT-H$k$-SJ MOSFET and the conventional SGT-MOSFET are shown in Fig. 3, showing the electric field profile along the right boundary line of the orange rectangular region in Fig. 1(c). The 3-D electric field distribution is presented to illustrate the electric field profile across the entire orange rectangular region in Fig. 1(c), a electric field drop under the Split-Gate region is demonstrated in Fig. 3(a) and 3(b). The vertical electric field analysis also demonstrates that incorporating a High-$k$ dielectric material located under the Split-Gate region significantly alters the electric field in the drift region. This results in the formation of new electric field peaks, with magnitudes reaching up to 0.275 MV/cm \cite{Investigationsof4H-SiCtrenchMOSFET}. Additionally, the 3-D electric field distribution shows that the BT-H$k$-SJ MOSFET structure provides a more uniform electric field profile, characterized by lower peak values compared to the conventional SGT-MOSFET. This improvement in the electric field distribution leads to a higher $\text{BV}$ for the device.  

\begin{figure}
    \centering

        \includegraphics[width=1\columnwidth]{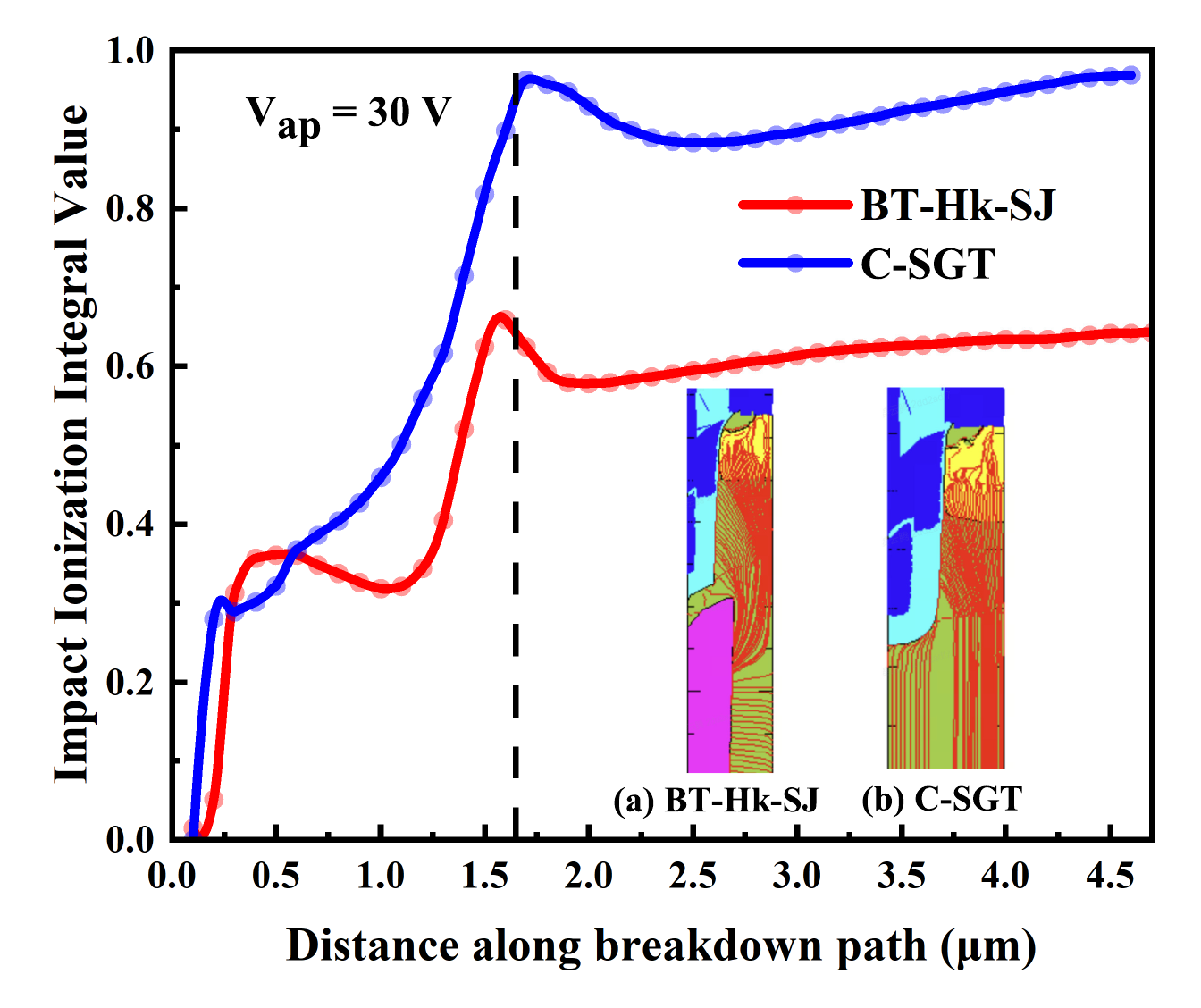}
        
    \hspace{0.01\linewidth} 
    \vspace{-6mm}
  
      \caption{Impact ionization integral and electric lines of BT-H$k$-SJ MOSFET and SGT-MOSFET}
    
\end{figure}

\par Fig. 4 illustrates the impact ionization integral and electric field lines of both the BT-H$k$-SJ MOSFET and conventional SGT structures. As shown in figure, the impact ionization integral of the BT-H$k$-SJ MOSFET remains consistently lower than that of the conventional SGT across almost all regions along the same path. Moreover, the peak impact ionization integral value of the BT-H$k$-SJ MOSFET is notably reduced, showing a decrease of over 32\% compared with the conventional SGT. Additionally, the electric field lines in BT-H$k$-SJ MOSFET exhibit a more uniform and sparser distribution with some electric field lines being absorbed into the High-$k$ region. The impact ionization integral profile and electric field distributions indicate an optimized voltage buffering ability within the device due to the new structural design, supporting the effectiveness of the High-$k$ SGT in enhancing device performance through breakdown voltage optimization.

\par The behavior observed highlights the importance of the High-$k$ material in modulating electric field characteristics, ultimately leading to improved performance metrics for power electronic devices.



\subsection{\textbf{Static Characteristic}}

\par The output characteristics of BT-H$k$-SJ MOSFET and SGT-MOSFET are depicted in Fig. 5(a), where $V_{\rm{GS}}$ are 4.5V, 6V and 10V. The comparison between the BT-H$k$-SJ MOSFET and the SGT-MOSFET is conducted under the same $V_{\rm{th}}$. However, the drain current $I_{\rm D}$ of the BT-H$k$-SJ MOSFET is significantly higher, primarily due to the modulation of the electric field by the integrated Bottom-Trench High-$k$ pillar and gate dielectric. This structure enhances the doping concentration, which in turn reduces the on-resistance and promotes improved forward conduction in the device \cite{Mingmin_2016} \cite{Investigationsof4H-SiCtrenchMOSFET}. The Fig.5 (b) illustrates the transfer characteristics of BT-H$k$-SJ MOSFET and SGT-MOSFET. The on-resistance of BT-H$k$-SJ MOSFET is reduced due to the presence of the High-$k$ Superjunction. This allows for the optimization of the gate dielectric thickness and the $p$-body concentration during the fabrication process, with minimal trade-offs. As a result, the transconductance of the new device is significantly improved.

\begin{figure}
    \centering
    \subfigure[]{
    \includegraphics[width=0.45\columnwidth]{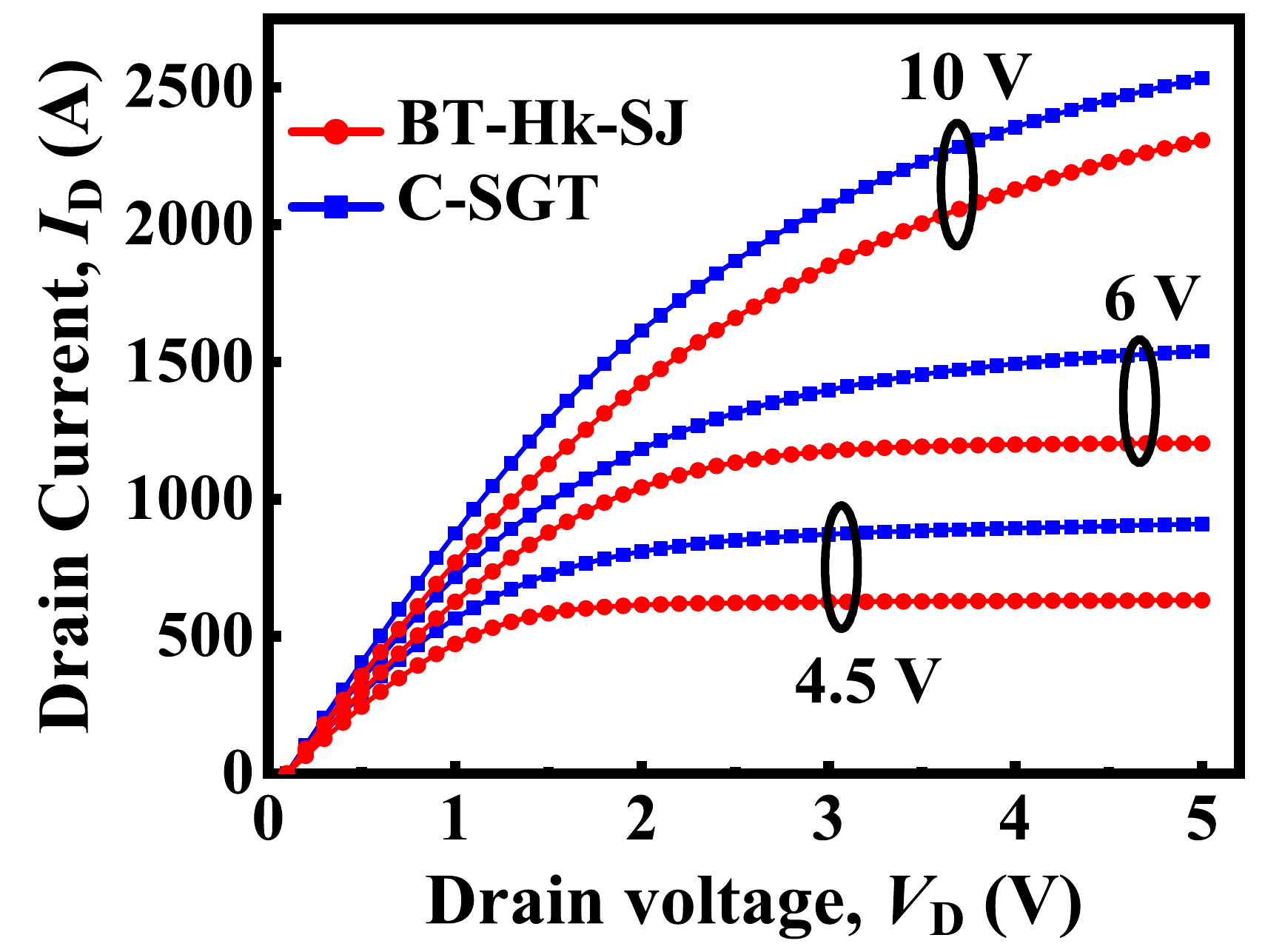}
        
    }\hspace{0.01\linewidth} 
    \subfigure[]{
        \includegraphics[width=0.46\columnwidth]{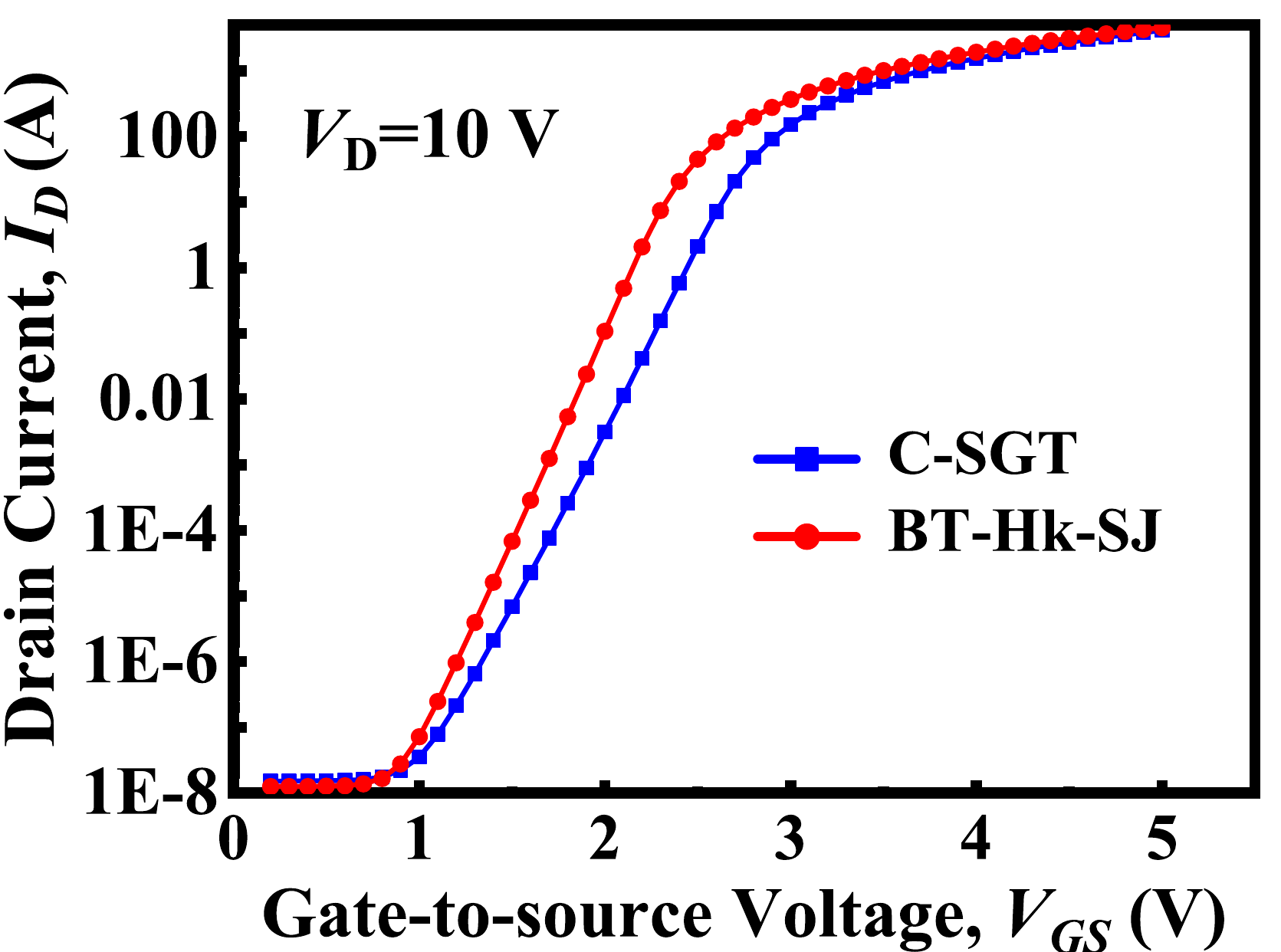}
        
    }\hspace{0.01\linewidth} 
       \vspace{-4mm}
       
    \caption{The I-V Characteristics of BT-H$k$-SJ MOSFET and SGT-MOSFET (a) output characteristics (b) transfer characteristics}
    
\end{figure}

\par The temperature dependence on {$V$\textsubscript{th}} and {$R$\textsubscript{DS}} are demonstrated in Fig. 6. Disregarding the work function difference between the gate polysilicon and the semiconductor, the {$V$\textsubscript{th}} is expressed as \cite{gu2024investigation}:

\begin{equation}
V_{{\mathrm{th}}}=\frac{\sqrt{4\varepsilon_{\mathrm{s}}N_{\mathrm{A}}q\psi_{\mathrm{B}}}}{C_{\mathrm{OX}}}+2\psi_{\mathrm{B}}-\frac{Q_{\mathrm{OX}}}{C_{\mathrm{OX}}} 
\end{equation}

\par From equation (1), {$\varepsilon$\textsubscript{s}} represents the dielectric constant of the semiconductor, {$N$\textsubscript{A}} denotes the doping concentration of the $p$-body, $q$ is the electron charge, {$C$\textsubscript{OX}} is the capacitance of the oxide layer, {$Q$\textsubscript{OX}} refers to the total effective charge in the oxide, {$\psi${\textsubscript{B}}} is the bulk potential of the semiconductor. The temperature-dependent parameters in equation (1) are primarily {$Q$\textsubscript{OX}} and {$\psi${\textsubscript{B}}}. Additionally, at low temperatures, the incomplete ionization of $p$-type must also be taken into account. {$\psi${\textsubscript{B}}} demonstrates a nearly linear negative temperature dependence over a broad temperature range \cite{du2020estimating}. The temperature variation of is primarily influenced by the interface state trap charge ({$Q$\textsubscript{Dit}}). At low temperatures, more electrons become trapped at the SiO$_2$/Si interface, leading to an increase in the magnitude of {$Q$\textsubscript{Dit}}. Since {$Q$\textsubscript{Dit}} is electronegative, this results in a reduction of at lower temperatures. Taking all these factors into consideration, {$V$\textsubscript{th}} exhibits a significant increase at low temperatures. The {$V$\textsubscript{th}} of BT-H$k$-SJ MOSFET and SGT-MOSFET exhibits a negative temperature dependence in Fig. 6. 

\par The temperature dependence of {$R$\textsubscript{DS}}, as shown in Fig. 6, does not follow the typical monotonic increase observed in standard trench MOSFET. Instead, it first decreases and then increases. The latter increase in on-resistance can be attributed to the reduction in mobility within the drift region and the inversion layer \cite{BaligaB.Jayant2008Fops}. The unusual behavior observed in the initial part of the curve is believed to result from the rise in {$V$\textsubscript{th}} with decreasing temperature. At this stage, the channel resistance ($R$\textsubscript{CH}) can be expressed as \cite{BaligaB.Jayant2008Fops}: 
\begin{equation}
    R_{\mathrm{CH}}=\frac{L_{\mathrm{CH}}}{Z\mu_{\mathrm{ni}}C_{\mathrm{OX}}(V_{\mathrm{G}}-V_{\mathrm{TH}})}
\end{equation}
\par Equation 2 demonstrates that {$R$\textsubscript{CH}} is inversely proportional to {$V$\textsubscript{th}}. Therefore, at low temperatures, the increase in {$R$\textsubscript{CH}} dominates, leading to the observed atypical relationship between {$R$\textsubscript{DS}} and temperature.

\begin{figure}
    \centering

        \includegraphics[width=1\columnwidth]{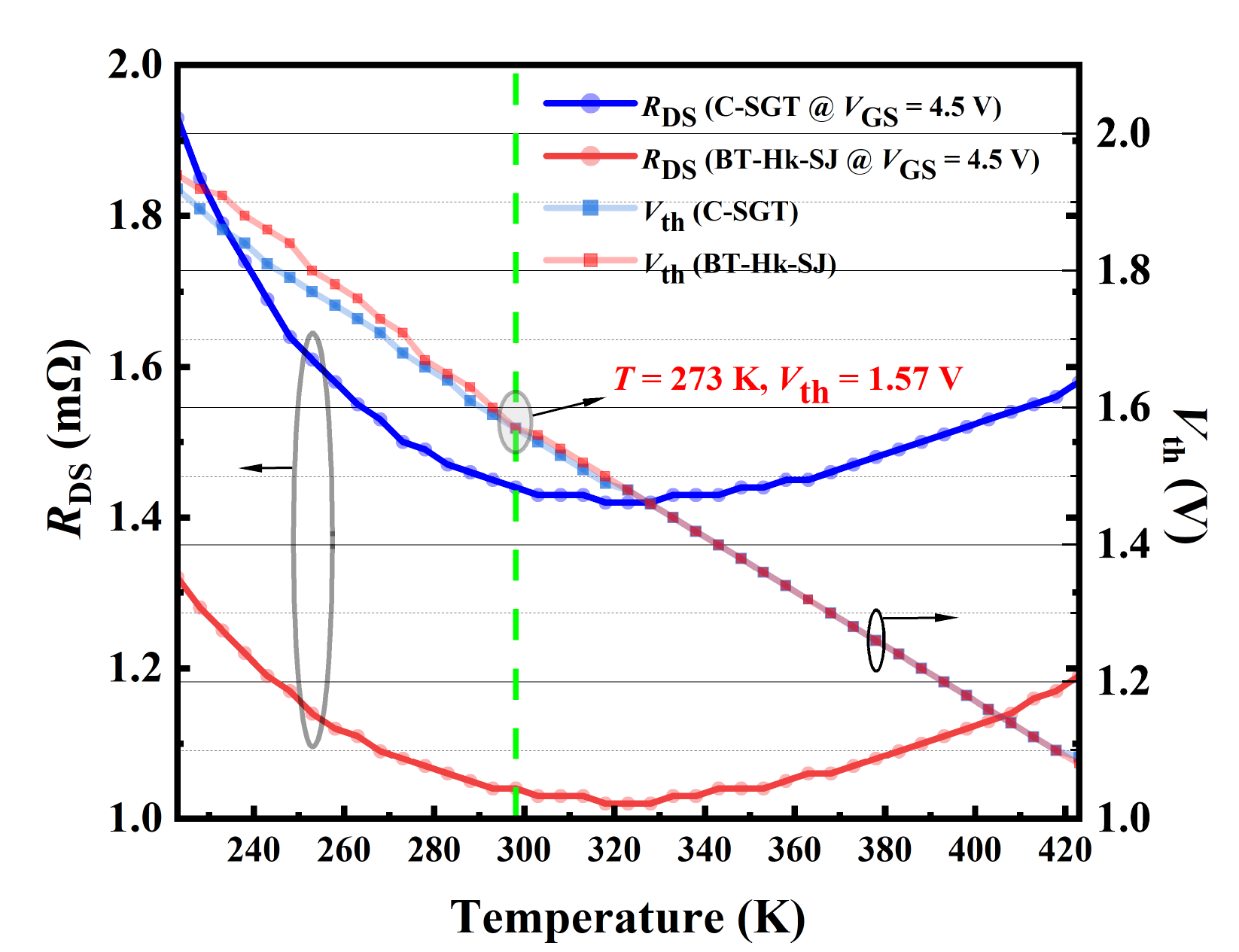}
       
    \hspace{0.01\linewidth} 
       \vspace{-6mm}
  
      \caption{Dependence of threshold voltage and drain-source resistance of BT-H$k$-SJ MOSFET and SGT-MOSFET on different temperature}
\end{figure}

\subsection{\textbf{Dynamic Characteristic}}

\par The result of input capacitance ({$C$\textsubscript{iss}}), output capacitance ({$C$\textsubscript{oss}}) and reverse transfer capacitance ({$C$\textsubscript{rss}}) with $V\textsubscript{DS}$ varies are presented in Fig.7. The capacitive characteristics of the BT-H$k$-SJ MOSFET show notable improvements over those of the conventional device, with enhancements of approximately 27\%, 21\%, and 24\% for {$C$\textsubscript{iss}}, {$C$\textsubscript{oss}} and {$C$\textsubscript{rss}}. These improvements can be attributed to the optimized structural design of the new device, particularly the integration of the High-$k$ Superjunction and the refined gate dielectric. These modifications facilitate better charge control, improved charge distribution, and a reduction in parasitic capacitance and overall greater device efficiency. Fig. 8 illustrates the switching delay time of two devices under identical conditions. The gate voltage pulse is applied with an amplitude of 10V, and both the rise time and fall time are set to 40 $ns$. As shown in the figure, the switching delay times of the BT-H$k$-SJ MOSFET and the SGT-MOSFET are nearly identical, indicating comparable switching performance between the two devices under the given operating parameters.

\begin{figure}[ht]
    \centering

        \includegraphics[height=6.4cm,width=0.9\columnwidth]{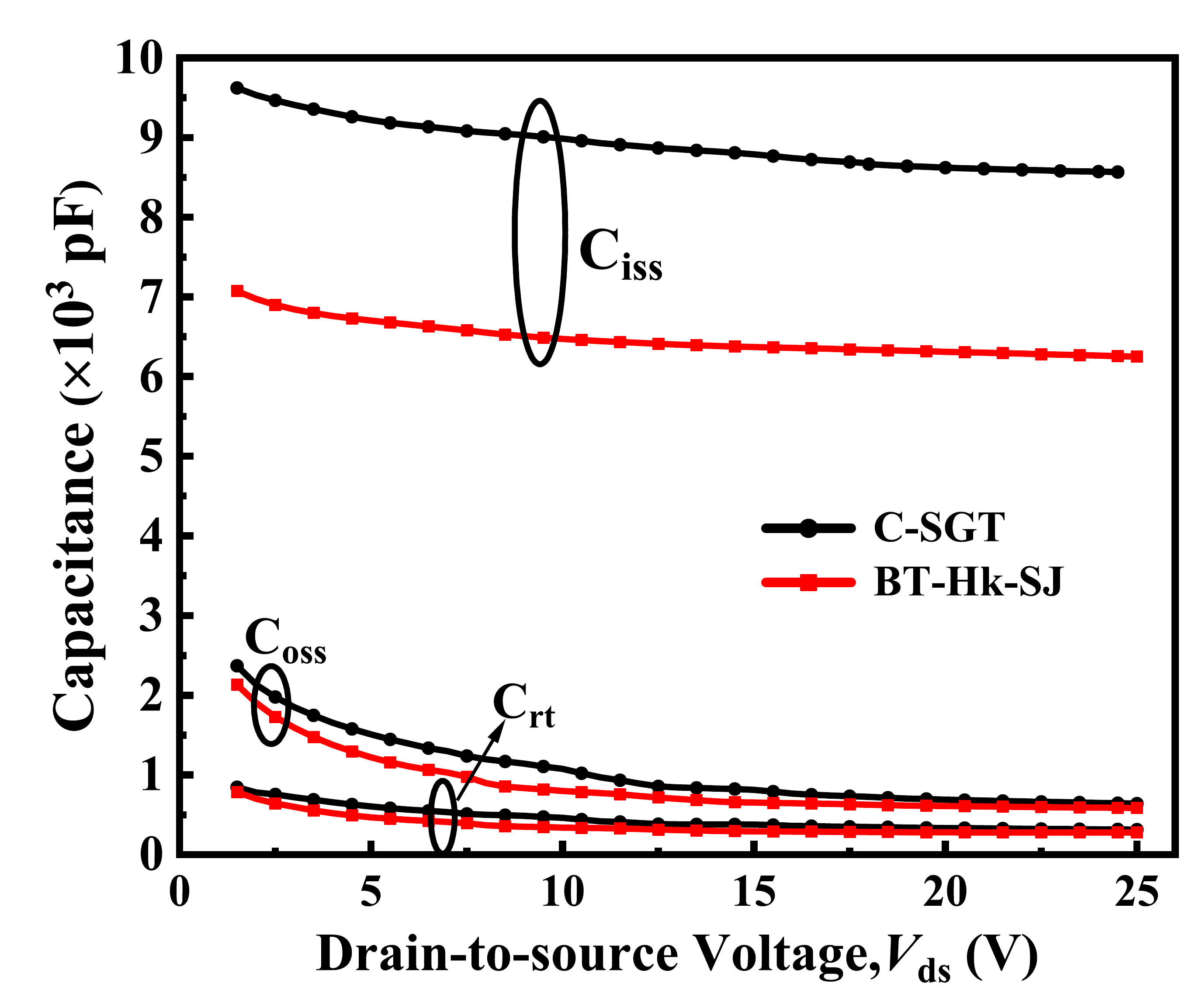}
        
    \hspace{0.01\linewidth} 
       \vspace{-4mm}
  
      \caption{MOSFET capacitance characteristics of BT-H$k$-SJ MOSFET and SGT-MOSFET}
\end{figure}

\begin{figure}[ht]
    \centering
     \includegraphics[height=6.4cm,width=0.9\columnwidth]{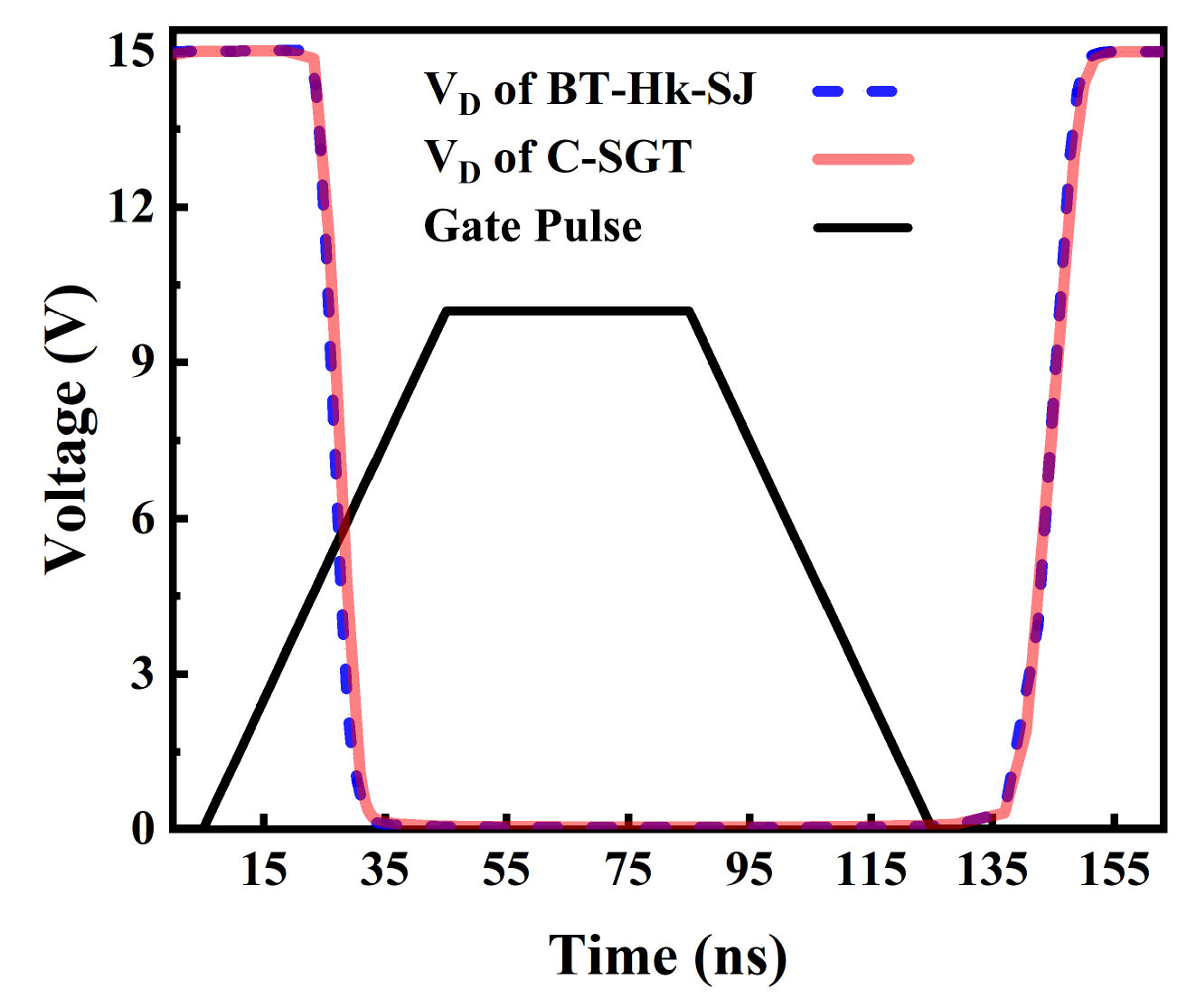}
        
    \hspace{0.01\linewidth} 
       \vspace{-4mm}
    \caption{Switching delay time of BT-H$k$-SJ MOSFET and SGT-MOSFET}
\end{figure}

\subsection{\textbf{Diode Characteristic}}
\begin{figure}[ht]
    \centering

        \includegraphics[height=6.4cm,width=0.94\columnwidth]{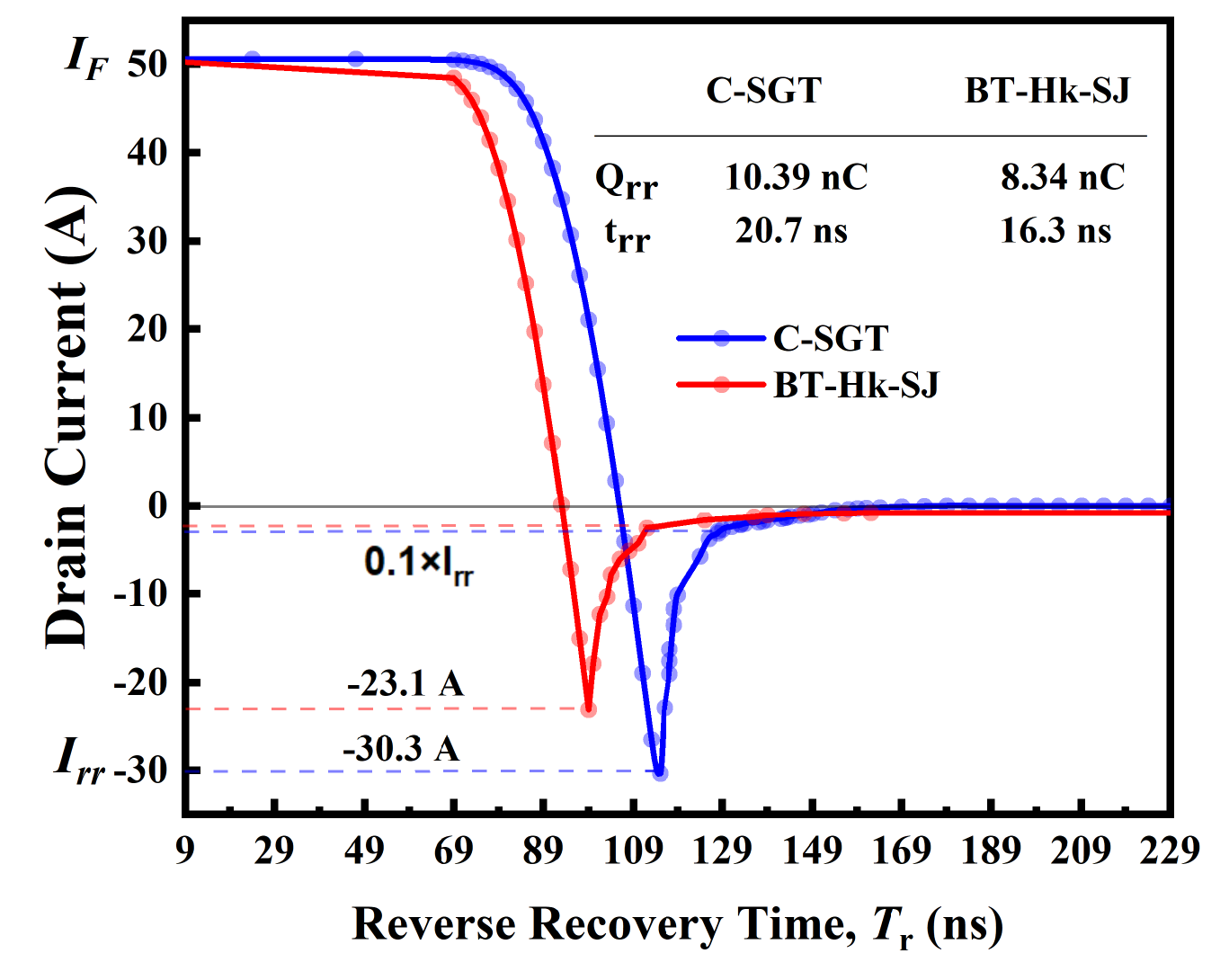}
    \hspace{0.01\linewidth} 
       \vspace{0mm}
  
      \caption{ Reverse recovery characteristics of BT-H$k$-SJ MOSFET and SGT-MOSFET}
\end{figure}

\par Reverse recovery characteristics of BT-H$k$-SJ MOSFET and SGT-MOSFET are demonstrated in Fig. 9. The incorporation of High-$k$ Superjunction significantly enhances the reverse recovery performance of the new device. Notably, the peak reverse recover current ({$I$\textsubscript{rr}}) of BT-H$k$-SJ MOSFET and SGT-MOSFET are 23.1A and 30.3A respectively , which is reduced by approximately 25\% \cite{kong2023novel}. More importantly, both the reverse recovery time ($T$\textsubscript{rr}) and reverse recovery charge ($Q$\textsubscript{rr}) are optimized by around 20\%. These improvements suggest that the new device will outperform conventional devices in applications involving shutdown and reverse blocking, offering superior efficiency and reliability in these operating conditions.

\subsection{\textbf{Analysis of FOMs}}

To effectively compare the performance of the conventional SGT-MOSFET and the optimized BT-H$k$-SJ MOSFET, key FOMs and device parameters are highlighted. Switching losses in MOSFETs arise from the finite switching time during turn-on and turn-off, alongside conduction losses due to $R_{\rm{DS,on}}$. The $Q_{\rm{GD}}\cdot R_{\rm{DS,on}}$ FOM is often used to balance switching and conduction losses, where a lower value indicates enhanced efficiency in high-performance applications like power converters, demanding both fast switching and low conduction losses \cite{FOM}.

For evaluating high-voltage performance, $\text{BV}^2/R_{\rm{on,sp}}$ is a classic FOM that combines voltage robustness with minimized conduction loss, making it ideal for assessing power density in high-voltage applications \cite{BaligaB.Jayant2008Fops, FOM, chengJEDS8759925}.

In high-frequency applications, Baliga’s High-Frequency Figure of Merit (BHFFOM)\cite{FOM} evaluates device efficiency by balancing $R_{\rm{on,sp}}$ (conduction losses) and $C_{\rm{in,sp}}$ (input capacitance-related switching losses). The BHFFOM is expressed as:

\begin{equation}
    \mathrm{BHFFOM}=\frac{1}{R_{\mathrm{on,sp}}\cdot C_{\mathrm{in,sp}}}
\end{equation}

For output switching-focused applications, the New High-Frequency Figure of Merit (NHFFOM)\cite{FOM,NHFFOM} substitutes \(C_{\text{out}}\) for \(C_{\text{in}}\), relevant for systems where output characteristics dominate:

\begin{equation}
    \mathrm{NHFFOM} = \frac{1}{{R_{\mathrm{on,sp}}} \cdot C_{\mathrm{out,sp}}}
\end{equation}

\begin{table}[ht]
  \centering
  \caption{\textbf{Comparison of FOMs between Conventional SGT-MOSFET and BT-H$k$-SJ MOSFET}}
  \resizebox{\columnwidth}{!}{ 
  \begin{tabular}{|c|c|c|c|c|}
    \hline
    \textbf{Device Structure} & \textbf{$Q_{\rm{GD}} \cdot R_{\rm{DS,on}}$} & \textbf{$\text{BV}^2/R_{\rm{on,sp}}$} & \textbf{BHFFOM} & \textbf{NHFFOM} \\ 
    \hline
    C-SGT & 1 & 1 & 1 & 1 \\ 
    \hline
    BT-H$k$-SJ & \textcolor{red}{0.9347} & \textcolor{red}{1.1582} & \textcolor{red}{1.3948} & \textcolor{red}{1.2857} \\ 
    \hline
  \end{tabular}
  }
\end{table}

The table II presents a comparison between our proposed BT-H$k$-SJ MOSFET and the conventional SGT-MOSFET, both configured with an identical threshold voltage ($V_{\text{th}}$) of 1.5V and comparable device dimensions. Each Figure of Merit (FOM) is normalized against the conventional SGT-MOSFET, which is set as the baseline (1) for comparative evaluation.

In particular, we observe the enhancements in key FOMs, such as $Q_{\text{GD}}\cdot R_{\text{DS,on}}$, $\text{BV}^2/R_{\rm{on,sp}}$, BHFFOM, and NHFFOM, where the BT-H$k$-SJ MOSFET demonstrates notable improvements. These values highlight the superior switching efficiency, higher breakdown resilience, and enhanced performance at high frequencies, establishing the BT-H$k$-SJ as a more efficient option in power and high-frequency applications compared to the standard SGT-MOSFET.

\section{Impact of Bottom-Trench High-$k$ Superjunction Pillar}

\subsection{\textbf{High-$k$ Pillar Depth}}

\begin{figure} 
    \centering

        \includegraphics[height=6.4cm,width=1\columnwidth]{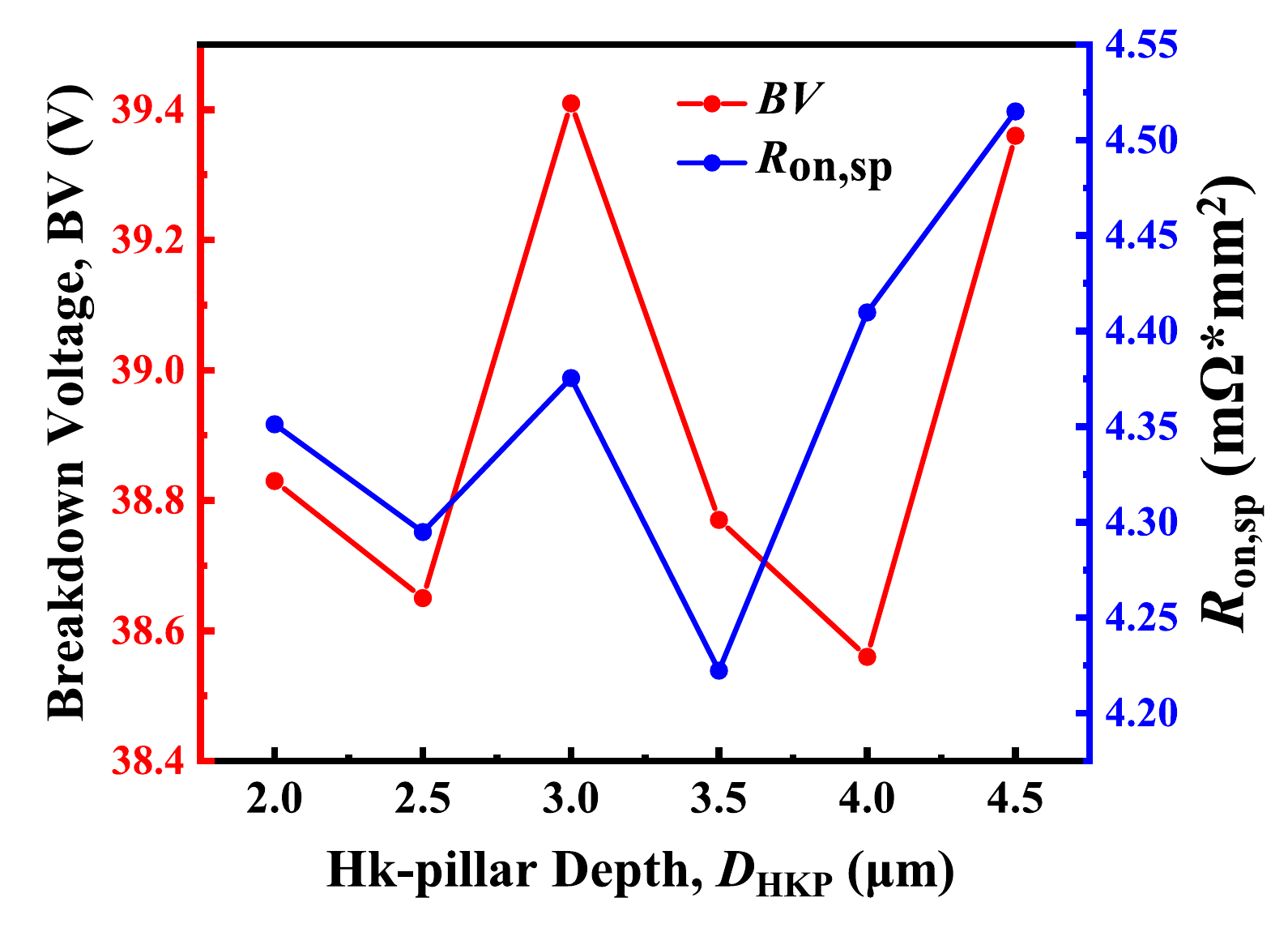}
    \hspace{0.01\linewidth} 
       \vspace{-8mm}
  
      \caption{Breakdown voltage and specific on resistance of BT-H$k$-SJ MOSFET at different High-$k$ pillar depth}
\end{figure}

\par Fig.10 illustrates the relationship between High-$k$ pillar depth ({$D$\textsubscript{HKP}}) with the breakdown voltage ($\text{BV}$) and sepcific on resistance ({$R$\textsubscript{on,sp}}) of BT-H$k$-SJ MOSFET. From the Fig. 10, it is evident that there is no consistent or regular relationship between $\text{BV}$ and  {$R$\textsubscript{on,sp}} with depth. Additionally, the line graph does not show a case where the {$R$\textsubscript{on,sp}} reaches its minimum value at the maximum $\text{BV}$. Therefore, in our optimization, a trade-off approach will be considered to determine the optimal value. Ultimately, a depth of 3 microns is chosen, as it provides a balance where the device $\text{BV}$ is maximized while the {$R$\textsubscript{on,sp}} remains relatively low.

\begin{figure}[htbp]
    \centering

        \includegraphics[height=6cm,width=0.9\columnwidth]{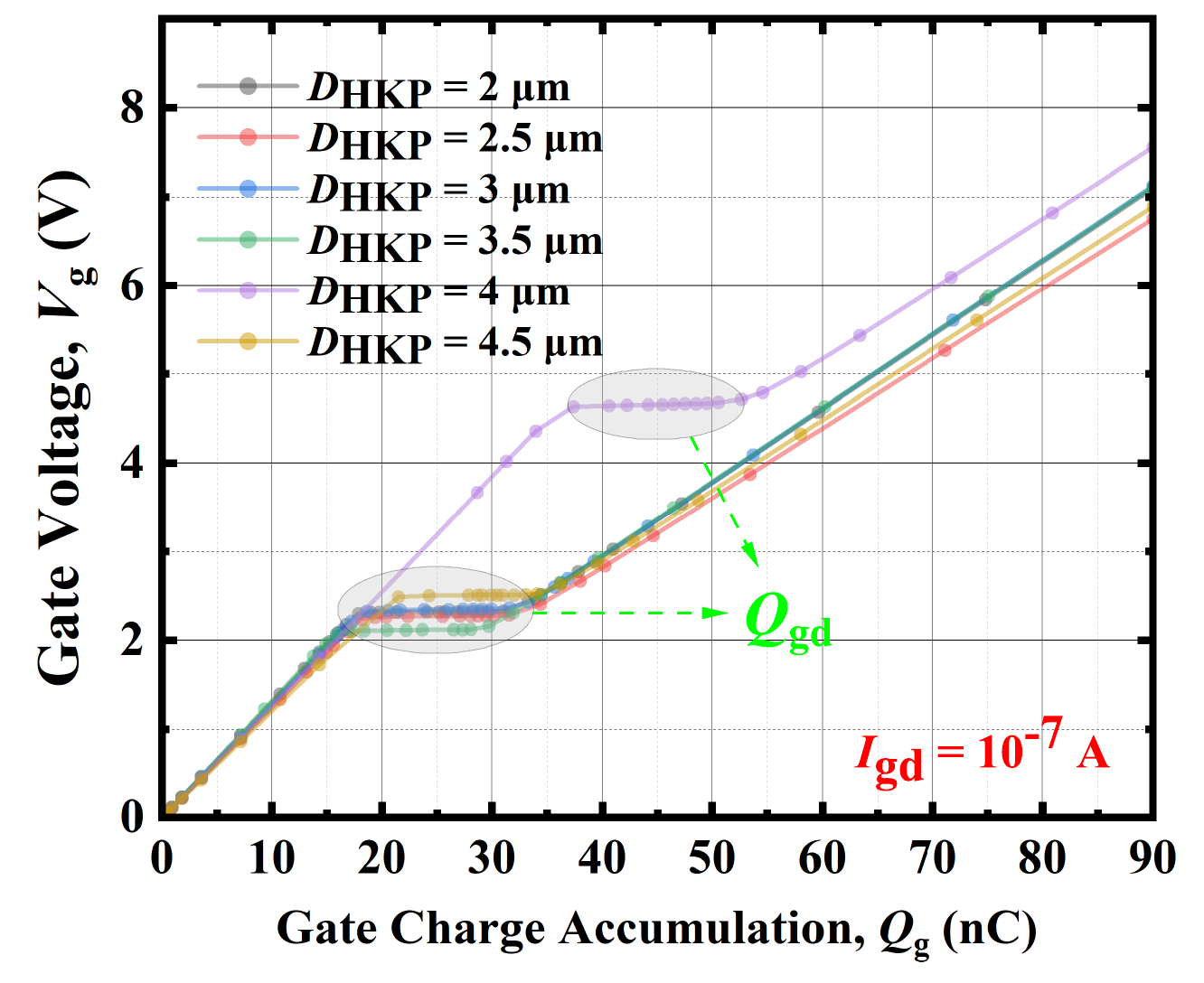}
    \hspace{0.01\linewidth} 
       \vspace{0mm}
  
      \caption{Gate voltage of BT-H$k$-SJ MOSFET during turn-on with constant gate current at different High-$k$ pillar depth  }
\end{figure}

\par The gate charge ($Q_{\text{G}}$) of BT-H$k$-SJ MOSFET varies with different High$k$ pillar depth is shown in Fig. 11. The gate-drain charge ($Q_{\text{GD}}$) is typically the dominant component of the switching charge and a key component of the FOM, making it a critical parameter in the design of power MOSFETs \cite{BaligaB.Jayant2008Fops} \cite{TongC.F.2009SaDA}. As observed in the figure, $Q_{\text{GD}}$ remains largely consistent across different configurations. However, the gate-source charge ($Q_{\text{GS}}$) and total gate charge exhibit more substantial variations. This variation is likely attributed to the changes in the shape and positioning of the split gate and the main gate as the pillar depth is adjusted during the fabrication process.

Table III provides a quantitative comparison of FOMs across various $D_{\rm{HKP}}$ values, from 2$\mu m$ to 4.5 $\mu m$ , highlighting the improvements achieved in breakdown voltage and switching characteristics with different pillar depths. The radar plot in Fig. 12 visually depicts the FOM performance across different depths, illustrating the trade-offs between FOMs. It reveals that the 3.5 $\mu m$  depth provides an optimal balance, achieving high breakdown voltage and minimal on-resistance without sacrificing other performance aspects. This analysis shows that the choice of High-$k$ pillar depth significantly influences the device's efficiency and reliability, with 3.5 $\mu m$  emerging as an ideal depth for achieving a balanced enhancement across all evaluated FOMs, thus supporting the design optimization strategy proposed in this study.

\begin{table}[!ht]
  \centering
  \caption{\textbf{Comparison of FOMs between Conventional SGT-MOSFET and BT-H$k$-SJ MOSFET}} 
  \resizebox{\columnwidth}{!}{ 
  \begin{tabular}{|c|c|c|c|c|}
    \hline
    \textbf{FOMs} & \textbf{$Q_{\rm{GD}} \cdot R_{\rm{DS,on}}$} & \textbf{$\text{BV}^2/R_{\rm{on,sp}}$} & \textbf{BHFFOM} & \textbf{NHFFOM} \\ 
    \hline
    C-SGT & 1 & 1 & 1 & 1 \\ 
    \hline
    2 $\mu$m & 0.7498 & 1.3367 & 1.3449 & 1.2590 \\ 
    \hline
    2.5 $\mu$m & \textcolor{red}{0.7172} & 1.3417 & 1.3091 & 1.2312 \\ 
    \hline
    3 $\mu$m & 0.7447 & 1.3694 & 1.3333 & \textcolor{red}{1.2505} \\ 
    \hline
    3.5 $\mu$m & 0.7369 & \textcolor{red}{1.3733} & \textcolor{red}{1.3948} & 1.2090 \\ 
    \hline
    4 $\mu$m & 0.8156 & 1.3007 & 1.3349 & 1.1932 \\ 
    \hline
    4.5 $\mu$m & 0.7946 & 1.3236 & 1.3130 & 1.1739 \\ 
    \hline
  \end{tabular}
  }
\end{table}

\begin{figure}[!ht]
    \centering

        \includegraphics[width=1\columnwidth]{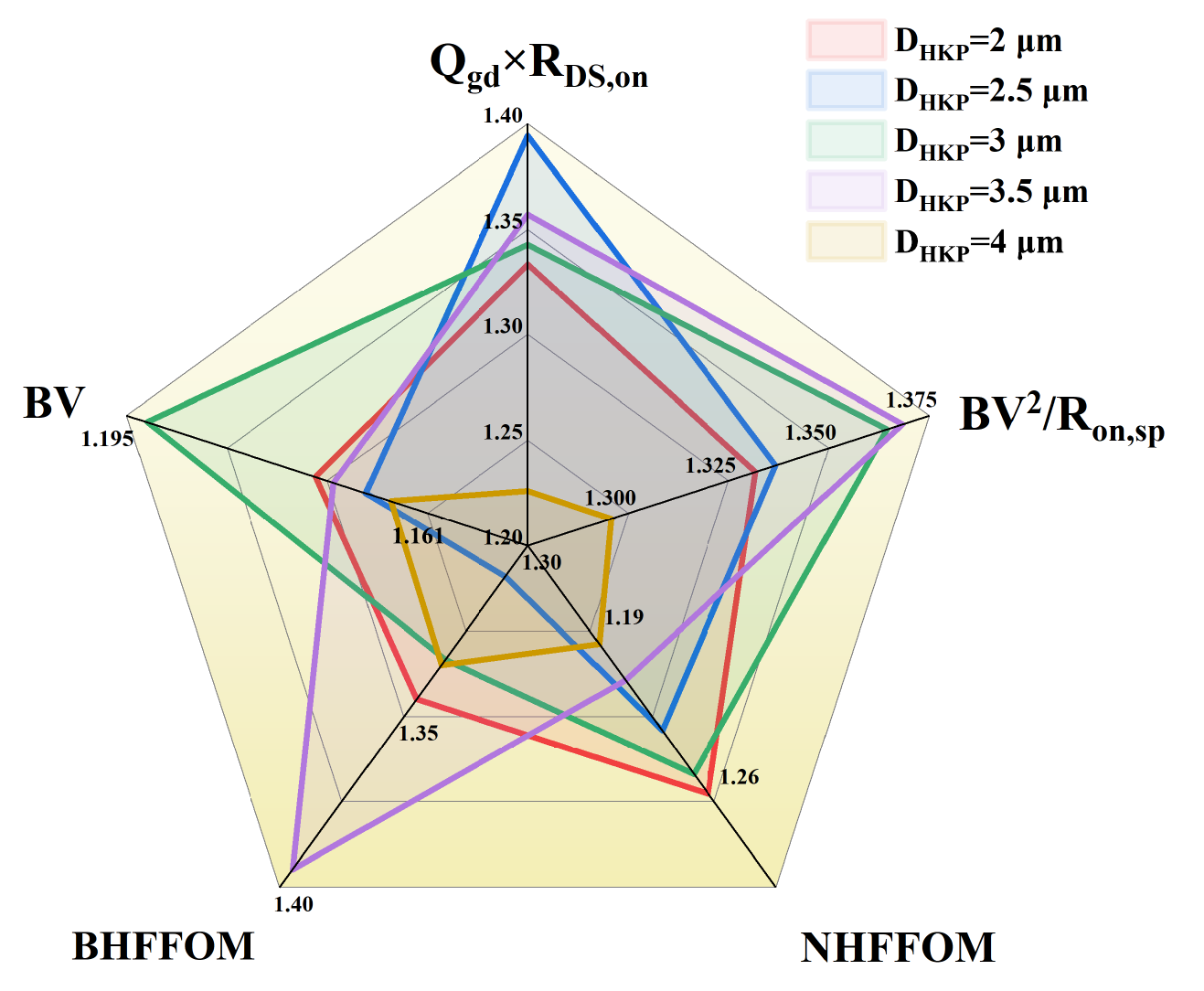}
    \hspace{0.01\linewidth} 
       \vspace{-6mm}
  
      \caption{Radar Map of FOM performance under different $D_{\rm{HKP}}$ }
\end{figure}

\par

\subsection{\textbf{High-$k$ Pillar Material}}

\begin{table}[htpb]
\centering
  \caption{\textbf{Comparison of Characteristics among different High-$k$ Pillar Materials }}
\begin{tabular}{|c|c|c|c|}
\hline
\diagbox{Characteristics}{Materials} & HfO$_{2}$                         & Nitride                      & Oxynitride                   \\ \hline
Drain-Source Resistance, $R_{\text{DS}}$ ($\Omega$)       & \textcolor{red}{ 0.433} &  0.437 &  0.492 \\ \hline
Breakdown Voltage, $\text{BV}$ ($V$)         & 35.14                        & 35.12                        & \textcolor{red}{ 35.74} \\ \hline
Gate-Drain Charge, $Q_{\text{GD}}$ ($nC$)        & \textcolor{red}{10.9}   & 11.3                         & 11.2                         \\ \hline
Reverse Recovery Charge, $Q_{\text{rr}}$ ($nC$)        & \textcolor{red}{ 8.34}  & 8.64                         & 8.7                          \\ \hline
\end{tabular}
\end{table}

\par This section examines the influence of strut materials on BT-H$k$-SJ MOSFET performance. As presented in Table IV, three materials that are well-established and feature simple fabrication processes hafnium dioxide (HfO$_{2}$), nitride, and oxynitride are compared. The table highlights the most representative parameters for static, dynamic, and diode characteristics, with red shading indicating the optimal values for each material. HfO$_{2}$ achieves the best performance across all three parameters. Although the oxynitride pillar demonstrates the highest breakdown voltage, HfO$_{2}$ exhibits the largest dielectric constant. This suggests that a material with an appropriate dielectric coefficient can contribute to a more uniform bulk electric field distribution within the structure \cite{cao2022tcad}. In general, the choice of material has a minimal impact on the device's electrical characteristics, with the variation across all parameters not exceeding 5\%.

\section{Conclusion}

This paper presents an optimized fabrication process for the Bottom-Trench High-$k$ Pillars Super-junction Split-Gate MOSFET, enhancing device performance with increased breakdown voltage, reduced on-resistance, and improved FOMs. The design balances performance and is cost-efficient, making it suitable for next-generation semiconductor applications. Simulations validated the High-$k$ pillar's role in electric field management, with 3.5 $\rm \mu m$ identified as the optimal depth. Analysis of High-$k$ materials showed  HfO$_2$'s superior performance due to its dielectric constant. The BT-H$k$-SJ MOSFET outperforms conventional designs, offering high efficiency and speed for power electronics, with a practical and scalable fabrication process for industrial use.


\bibliographystyle{IEEEtran}  
\bibliography{main}           

\end{document}